# COLOSSAL SPIN-PHONON ANOMALIES AND THE FERROELECTRIC PHASE TRANSITION IN THE MODEL MULTIFERROIC $BIFEO_3$

R.Haumont [1], J. Kreisel [1,*], P. Bouvier [2], F. Hippert [1]

[1] Laboratoire des Matériaux et du Génie Physique (CNRS), ENS de Physique de Grenoble, B.P.46, 38402 St. Martin d'Hères, France

[2] Laboratoire d'Electrochimie et de Physicochimie des Matériaux et des Interfaces (CNRS), ENSEEG, BP 75, 38402 St. Martin d'Hères Cedex, France

[*] Author to whom correspondence should be addressed. Email address: jens.kreisel@inpg.fr

**ABSTRACT**

We report a temperature-dependent Raman and neutron scattering investigation of the multiferroic material bismuth ferrite $BiFeO_3$ (BFO). The observed loss of the Raman spectrum at the ferroelectric Curie temperature $T_C$ gives evidence for a cubic $Pm\bar{3}m$ structure of the high-temperature paraelectric phase. The ferroelectric-to-paraelectric phase transition is not soft-mode-driven, indicating a non-conventional ferroelectric. Furthermore, our results reveal colossal phonon-anomalies around the magnetic Néel temperature $T_N$. We attribute these anomalies to the coupling of magnetic and ferroelectric order parameters. Our results point to Raman scattering as a powerful tool for the study of the coupling between magnetic and lattice degrees of freedom in multiferroics.

In past years there has been increasing interest in so-called magnetoelectric multiferroics, which show spontaneous magnetic and ferroelectric ordering, thus two ferroic properties, within the same phase. On the way towards a fundamental understanding of multiferroics, the experimental observation and understanding of the coupling mechanism between the ferroelectric and magnetic order are of great interest. Important advances were for instance made through the use of second harmonic light, dielectric and magnetic measurements or structural investigation.[1,2] In contrast, very little is known about the behavior of phonons in magnetoelectric multiferroics, and this although investigations of phonons have in the past played a crucial role in the understanding of classic ferroelectrics. Phonons are also known to be influenced by spin correlation thus offering a complementary tool.[3]

Recent investigations of $HoMnO_3$ by Raman and IR spectroscopy[4] and of $LuMnO_3$ by transmittance and reflectance measurements[3] have revealed the importance of phonon effects in multiferroics. The latter has been underpinned by both the observation of extraordinary spin-phonon interactions in $YMnO_3$ detected by thermal conductivity measurements[5] and the report of a strong spin-lattice coupling in $HoMnO_3$ derived from thermal expansion measurements.[6] A similar coupling is expected to exist in most, if not, all multiferroics but up to now the available data is restricted to rare earth (*RE)*-manganites.

Motivated to determine and understand the role of phonons in multiferroics, we have undertaken a first-time Raman scattering study of $BiFeO_3$ (BFO). Bismuth ferrite is an interesting model system because it presents a number of distinct features compared to *RE*-manganites: $BiFeO_3$ is a very robust multiferroelectric since it presents a coexistence of ferroelectric and antiferromagnetic order up to unusually high temperatures: In bulk single crystals, $BiFeO_3$ has an antiferromagnet (AFM) Néel temperature $T_N$ of ~370°C and a ferroelectric Curie temperature $T_C$ of ~830°C.[7] Furthermore, BFO presents exchange interactions only in the $Fe^{3+}$-subsystem, whereas *RE*-manganites present two correlated magnet subsystems ($RE^{3+}$ and $Mn^{3+}$). Moreover, the ferroelectricity in BFO is conditioned by the stereo-chemically active $6s^2$ lone pair of $Bi^{3+}$, while it is has been proposed that the ferroelectricity in *RE*-manganites originates in magnetic interactions.[2]

Our temperature-dependent Raman study reveals (*i*) The loss of the Raman spectrum at the Curie temperature $T_C$ indicating a cubic $Pm\bar{3}m$ structure for the paraelectric phase; interestingly this transition is not soft-mode-driven. (*ii*) Colossal phonon anomalies around the Néel temperature $T_N$ for some particular phonons, indicating a very strong spin-phonon coupling which we attribute to interactions between magnetic and ferroelectric order parameters. The reported anomalies demonstrate that phonon spectroscopy is a powerful tool for the study of such coupling in multiferroics.

Single crystals of $BiFeO_3$ were grown using a $Fe_2O_3/Bi_2O_3$ flux in a platinum crucible. Translucent crystals with a shape of thin platelets (≈1mm x 0.2 mm) have been isolated and Laue back-scattering indicates a $[001]_{pc}$ orientation of the platelets (pseudo-cubic setting). Raman spectra were recorded in back-scattering geometry with a LABRAM Jobin-Yvon spectrometer using a He-Ne laser (632.8 nm) as excitation line. Temperature measurements in the range from 25°C to 850°C have been carried out by using a commercial LINKAM heating stage placed under the Raman microscope. We have verified that the used laser power does not produce significant heating or damage the sample. Moreover, the Raman spectra before and after heating are identical, attesting the reversibility of temperature-induced changes. Contrary to earlier studies[8] we observe no sample decomposition above 810°C. Complementary neutron powder diffraction experiments ($\lambda$ = 2.52 Å) have been carried at station D1B at the ILL neutron reactor in Grenoble.

The room temperature structure of $BiFeO_3$ crystals is a highly rhombohedrally distorted perovskite with space group $R3c$.[9,10] The crystal structure of the high-temperature paraelectric phase has not been conclusively determined, a cubic $Pm\bar{3}m$ and a rhombohedral $R\bar{3}c$ phase being currently considered.[11] With respect to the cubic $Pm\bar{3}m$ structure the rhombohedral structure is obtained by an anti-phase tilt of the adjacent $FeO_6$ octahedra and a displacement of the $Fe^{3+}$ and $Bi^{3+}$ cations from their centrosymmetric position along $[111]_{pc}$. The 10 atoms in the unit cell of the rhombohedral *R*3*c* structure give rise to 15 Raman-active modes:

$$\Gamma_{Raman,R3c} = 4A_1 + 9E$$

Figure 1 presents Raman spectra of $BiFeO_3$ from room temperature up to 820°C. The Raman spectra are well-defined and provide reference spectra for both thin film investigations and a potential countercheck of first principles ab-initio calculations.[11,12] The overall spectral signature does not change notably in the range from 25°C - 800°C. This temperature behavior indicates that $BiFeO_3$ maintains its room temperature structure up to 800°C, which is in agreement with previous investigations.[8,9] With increasing temperature all bands shift to low-wavenumber (Figure 2) and broaden, a behavior which is explained by thermal expansion and thermal disorder, respectively.

While the Raman response is still very well-defined at 800°C, it is suddenly strongly reduced at 810°C and, finally, totally lost at 820°C (Figure 1). Such a spectral evolution is clear evidence for a structural phase transition towards a high-temperature cubic $Pm\bar{3}m$ phase for which any first-order Raman scattering is forbidden. The latter observation is not unusual knowing that most perovskite-type oxides crystallize at high-temperature in a $Pm\bar{3}m$ structure. Note that the ferroelectric (*R*3*c*)-to-paraelectric ($Pm\bar{3}m$) phase transition is *not* accompanied by the observation of a soft mode. At first sight, this is unexpected knowing that the ferroelectric-paraelectric phase transition of classic ferroelectrics like $PbTiO_3$ is soft-mode-driven. However, the absence of a soft mode can be related to the fact that the space groups *R*3*c* and $Pm\bar{3}m$ are not in a pure super- or sub-group relation, implying a first-order (not soft mode-driven) transition as illustrated by the abrupt loss of the Raman signature.

We now discuss the evolution of the Raman signature around the antiferromagnetic-paramagnetic (AFM-PM) phase transition ($T_N \approx 370$ K)[7]. Let us first recall that this magnetic transition is not accompanied by a structural phase transition[8,9], as supported by the fact that the overall raman spectral signature is maintained across the AFM-PM. Nevertheless, a closer inspection of the temperature-dependent Raman spectra reveals noticeable spectral changes in the vicinity of $T_N$. Upon cooling, the most remarkable observation is a colossal continuous step-like anomaly in the evolution of wavenumber for several bands (Figure 2). To place emphasis on these anomalies, figures 3a and 3b present a more detailed view for some bands. We note that the earlier reported[3,4,6] phonon anomalies for the multiferroics $HoMnO_3$ and $LuMnO_3$ are also large and that the spin-phonon interactions in the multiferroics $YMnO_3$ are considered to be extraordinary.[5] However, the reported *outstanding* anomalies[3,4] in the phonon spectra of the multiferroics $HoMnO_3$ and $LuMnO_3$ are by a factor of ~ 3 to 5 smaller than our observed *colossal* anomalies in BFO (Figure 3c). We also observe anomalies in the full width at half maximum (*FWHM*) and in intensity, again clearly pointing at two changes of regime (Figure 3d). The occurrence of such a change of signature can be understood within the concept of hard mode spectroscopy. This concept predicts that any change in structural and/or physical properties will, in principle, lead to variations in all phonon characteristics (frequency, intensity and *FWHM*).[13]

At $T_N$, the anomalies in wavenumber are characterized by a sudden frequency hardening : e.g. the band at 380 $cm^{-1}$ crosses upon cooling from a rate of $\Delta\omega/\Delta T \approx$ - 0.02 $cm^{-1}$/°C in the PM-phase (region I) to a pronounced hardening of ≈ - 0.2 $cm^{-1}$/°C in the AFM-phase (region II). Upon further cooling, the $\Delta\omega/\Delta T$ slope reverts back to a slow rate of ≈ - 0.015 $cm^{-1}$/°C (region III) at a temperature hereafter called $T^* \approx 275$°C. The low-wavenumber shift in regions I and III corresponds to a classic Grüneisen-type anharmonicity-related wavenumber hardening. On the other hand, the application of the Grüneisen formula $\Delta\omega/\omega = \gamma\ \Delta a/a$ with available X-ray diffraction data[8] to the intermediate region II leads to a Grüneisen parameter of $\gamma \approx 30$, indicating an important anharmonicity which cannot be explained by a classical temperature behavior but points at a further contribution.

At first sight, the considerable spectral changes at $T^*$ suggest a structural rearrangement which has to be subtle or to occur on a local level since they are not detected by X-ray diffraction.

Although we cannot formally exclude a change in space group at $T^*$, such a scenario is unlikely due the absence of any new spectral signature (i.e. new bands, band splitting etc.). Literature work allows ruling out a further magnetic phase transition and we have neither evidence nor reason to consider an electronic rearrangement.

A possible source of the observed anomaly at $T_N$ is magnetostriction. Diffraction experiments on $BiFeO_3$ show upon cooling indeed a sudden change (increase) in the lattice parameters at $T_N$ (Figure 3a).[8] However, the observed change in lattice parameter is by far too small to cause the observed phonon anomaly and, moreover, an increase in volume should rather lead to a decrease in wavenumber.

The observed phonon anomaly near $T_N$ for BFO is reminiscent of similar (though smaller) observations near magnetic phase transition in other oxides. For some of them, $SrRuO_3$ (ref. [14]), $A_2Mn_2O_7$ ($A$ = Tl, In, Y)[15] or $La_{0.7}Ca_{0.3}MnO_3$ (ref. [16]), the observed anomaly has been attributed to a strong electron-phonon coupling where the free carriers (or polarons) contribute in an important manner to the effective force constant. We do not expect that a ferroelectric insulator such as $BiFeO_3$ is described by the same physics.

On the other hand, the observed coupling between spin and phonon degrees of freedom in systems like $MF_2$ ($M$ = Fe, Mn)[17], $Y_2Ru_2O_7$ (ref. [18]), $LaTiO_3$ (ref. [19]) or $ZnCr_2O_4$ (ref. [20]) is closer to our observation. The qualitative behavior of the above materials is similar to what is observed for BFO, but the change in wavenumber observed for BFO is significantly larger. A useful approach for the understanding of spin-dependent phonon frequencies of the latter and other materials is based on an initial model by Baltensperger and Helmann, which considers the modulation of the magnetic exchange integral by the ionic displacement of the involved phonon mode.[21] Based on this approach, it has been proposed that the phonon frequencies in magnetic materials are affected by the correlation of spins of nearest-neighbor pairs.[17,21] In the case of multiferroics we might, however, expect a further coupling to the ferroelectric (and/or ferroelastic) order parameters. A magneto-electric coupling in terms of Landau theory has been discussed by Smolenskii and Chupis[22] who expressed the thermodynamic potential $F$ as following:

$$F = F_0 + \alpha P^2 + \frac{\beta}{2} P^4 - PE + \alpha' M^2 + \frac{\beta'}{2} M^4 - MH + \gamma P^2 M^2 + \ldots \qquad (1)$$

, where $P$ and $M$ are the polarization and the magnetization, respectively. The coupling between magnetic and ferroelectric order parameters is described by the term $\gamma P^2 M^2$. For $BiFeO_3$, our work shows that equation *(1)* should be extended by an additional $\frac{\gamma}{6} P^6$ term, which takes into account the first-order nature of the ferroelectric-to-paraelectric phase transition.

Kimura *et al*[23] have analysed equation *(1)* for the multiferroic $BiMnO_3$ and have proposed that the temperature dependence of the electric order parameters can be neglected at the magnetic transition, provided that the magnetic and ferroelectric phase transitions are separated apart. This condition is well respected for $BiMnO_3$ ($T_{FM}$ = -160 °C, $T_C$ = 480 °C) and for this material the authors show that the observed physical properties below $T_{FM}$ are proportional to the square of the magnetic order parameter.

Motivated to find out if such a scaling can also describe the physics of BFO, we have investigated a BFO powder sample by neutron diffraction in the range from ambient temperature to 420 °C. Figure 4 shows the evolution with temperature of the intensity for the $(101)_h$ magnetic Bragg reflection which is considered to be proportional to the magnetic order parameter. The plot of the squared intensity against temperature displays two changes of regime: A first change at $T_N$, as expected, and a second occurring interestingly at $T^*$, a temperature which was before observed by Raman scattering. $T^*$ is evidenced by the deviation from the linear regime in figure 4. It is just this correspondence between Raman scattering and magnetic diffraction, which further highlights the occurrence of a spin-phonon coupling.

It is natural to link the observed colossal coupling of BFO to its multiferroic character. This outstanding coupling is most possibly due to the fact that the magnetic and ferroelectric order temperatures in BFO are closer together when compared to the well-separated temperatures for multiferroics such as $HoMnO_3$ and $LuMnO_3$. This leads for BFO to a situation where the temperature dependence of the ferroelectric order parameter can no more be neglected at $T_N$, and thus the magnetic phase transition takes place within a phase that presents lattice instabilities. Such ferroelectric (cation displacement) and ferroelastic (octahedron tilt) instabilities are known to be very sensitive to external perturbations such as temperature, pressure, or strain. In a similar way, local spin correlations can act as a perturbation.

A possible scenario is the following: Upon cooling and while entering the magnetic phase at $T_N$, the atoms start to move according to their structural instabilities. This leads to the observed marked $\Delta\omega/\Delta T$ slope and an enhanced spin-phonon coupling for those phonons which involve a modulation of the Fe-O-Fe bond angle, the critical parameter in the magnetic superexchange. However, at a certain temperature, which turns out to be $T^* \approx 275$ °C, the lattice instabilities slow down and the atoms settle eventually into a new equilibrium position, from whereon ($T < T^*$, region III) the evolution can again be described with a common temperature-dependent anharmonic behavior.

More experimental and theoretical work is needed, however, to unambiguously identify the physical mechanisms leading to the anomalies reported here. First principle calculations of the phonon spectrum with models including the magnetic superexchange interactions as well as lattice instabilities could lead to a better understanding of the magnetoelectric coupling.

*The authors would like to thank ILL for access to their neutron facility and O. Isnard for help with the experimental setup.*

**FIGURES CAPTIONS**

FIGURE 1. Temperature-dependent Raman spectra of $BiFeO_3$.

FIGURE 2. Temperature-dependent evolution of the Raman band position for $BiFeO_3$.

FIGURE 3. (color on line) Temperature-dependent evolution of some spectral characteristics for $BiFeO_3$.

a) wavenumber-shift for the 380 $cm^{-1}$ band (left), change of hexagonal cell volume (right)[8].

b) wavenumber-shift for the 315 and 550 $cm^{-1}$ bands.

c) $\Delta\omega$ against temperature for the 380 $cm^{-1}$ band, $\Delta\omega$ being defined as $\omega - \omega_I$ after subtraction of the linear behaviour in regime I.

d) FWHM for the 380 $cm^{-1}$ band,

e) FWHM for the 315 and 550 $cm^{-1}$ bands.

FIGURE 4. Temperature-dependent evolution of the squared intensity of the magnetic Bragg reflection $(101)_h$. The dashed line is a guide to the eyes and corresponds to a linear fit of the low temperature regime.

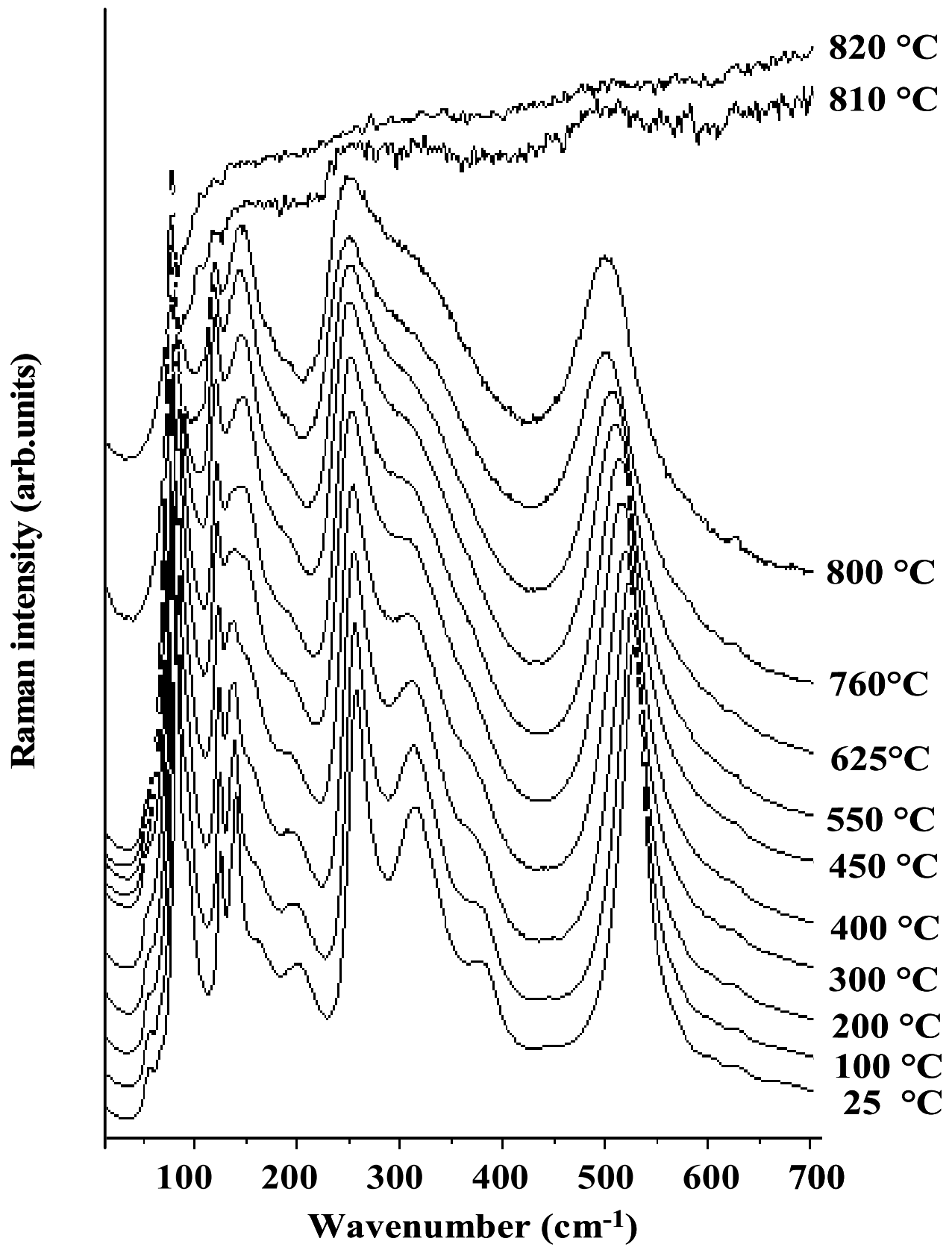

FIGURE 1

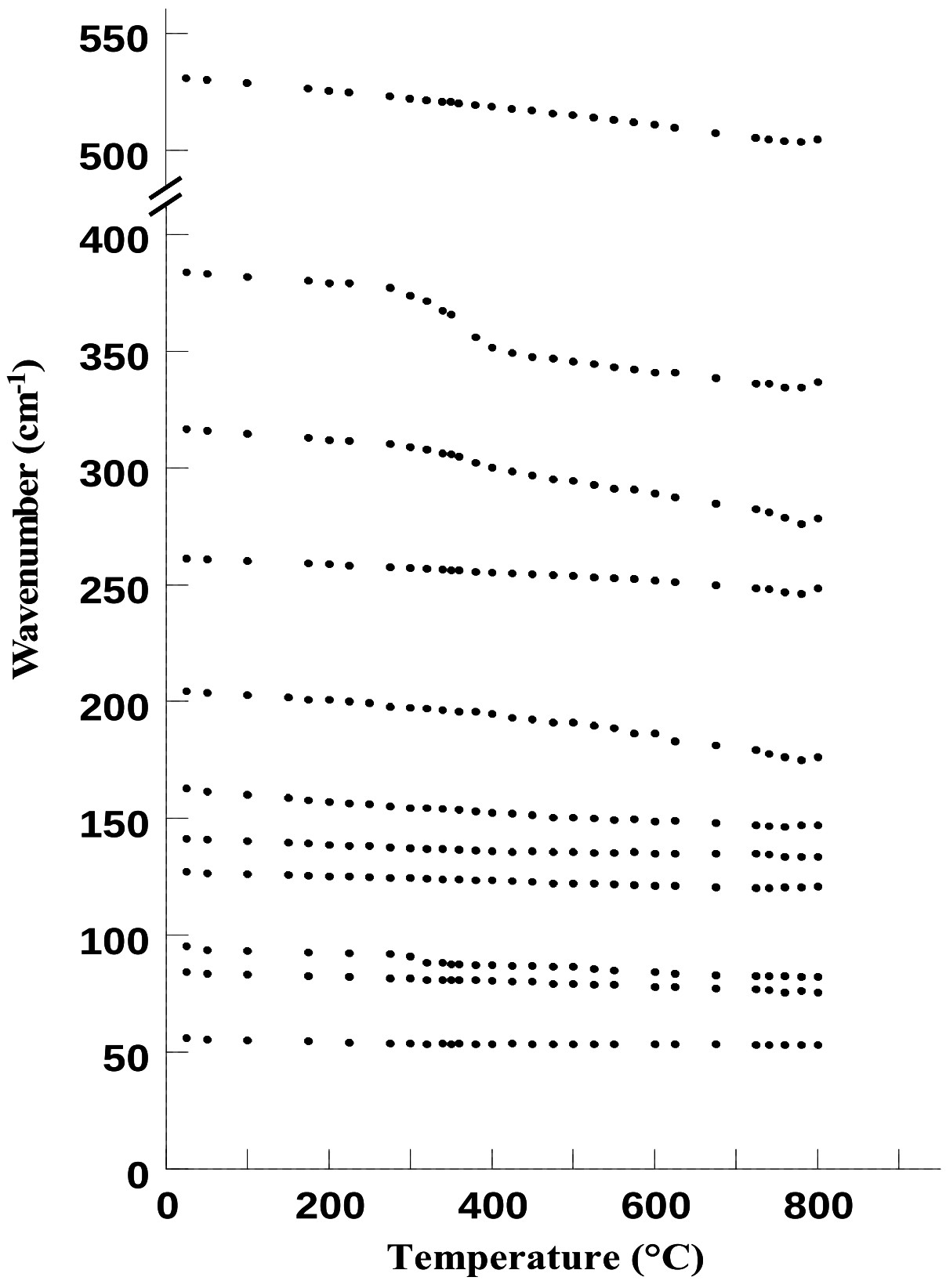

FIGURE 2

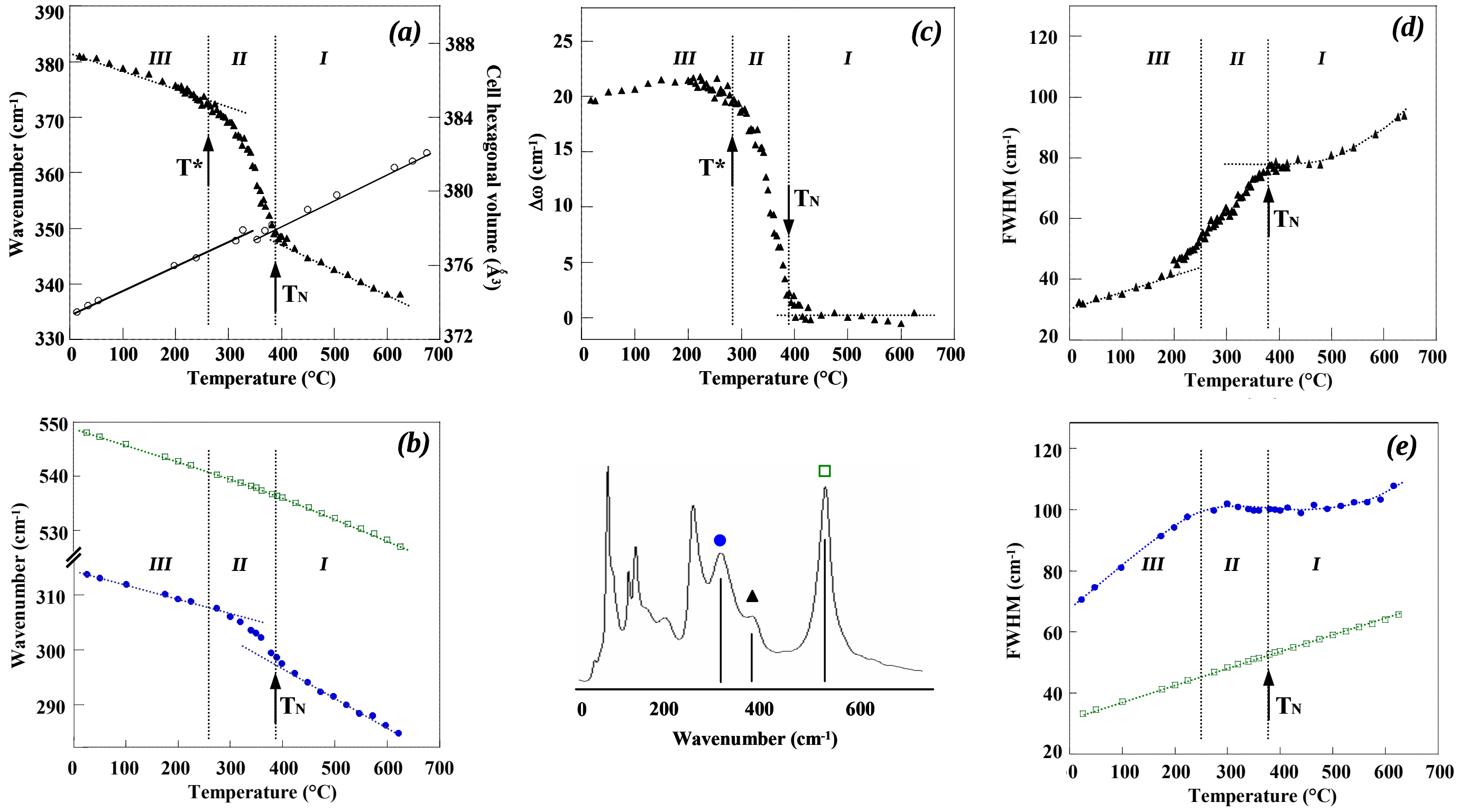

FIGURE 3

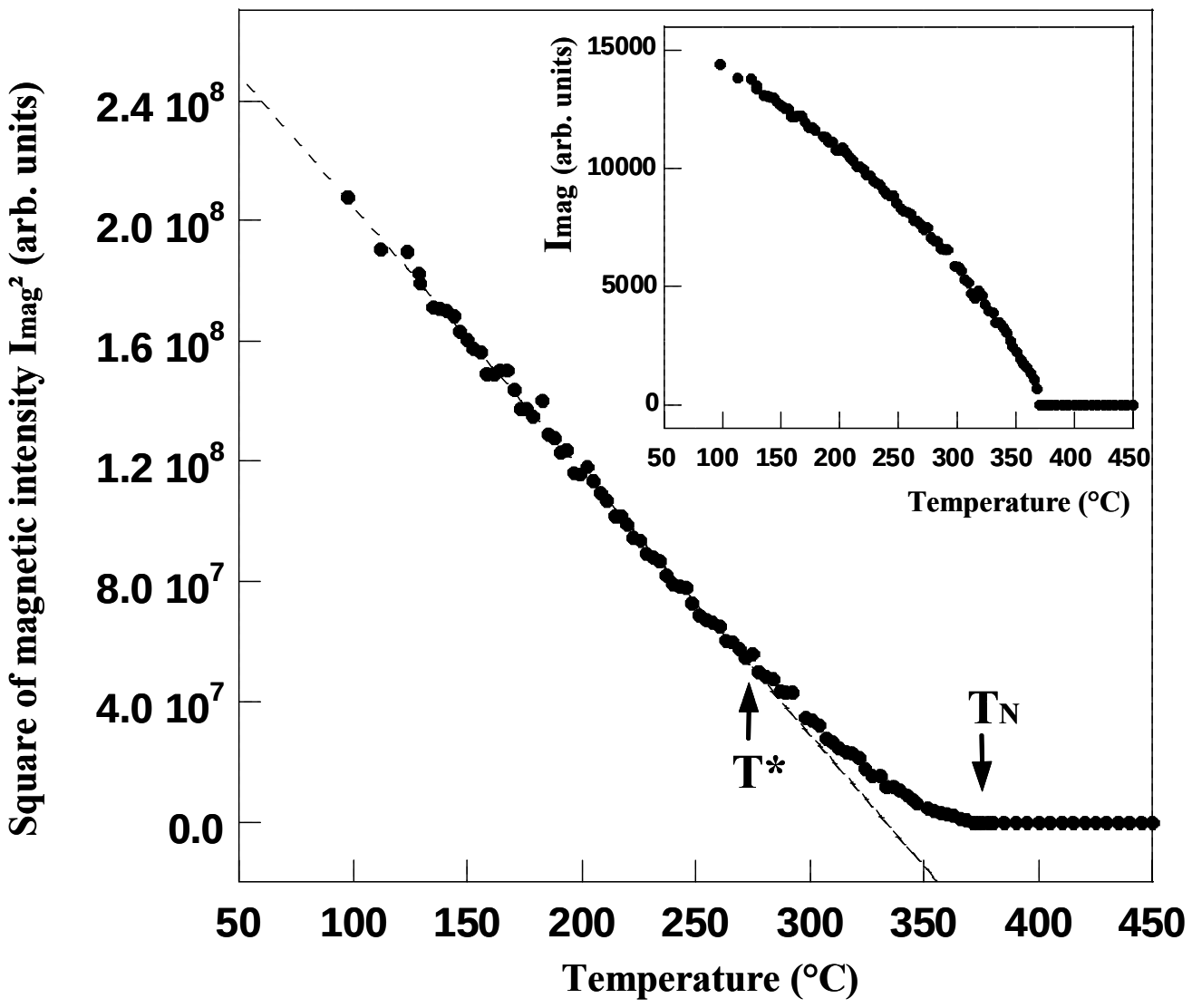

FIGURE 4